\documentclass[showpacs,twocolumn,prd,aps]{revtex4}
\usepackage{amsfonts}
\usepackage{amsmath}
\usepackage{amsbsy}
\usepackage{amssymb}
\usepackage{eepic}
\usepackage{rotating}

\font\grb=eurb10

\def\bphi{\hbox{\grb\char'047}\,}
\def\bpsi{\hbox{\grb\char'040}\,}

\def\vpint{\mathop{\scriptstyle{\mathbf{\diagup}}\hskip-2.2ex \displaystyle{\int}}}

\begin{document}

\title{Monodromy transform and the integral equation method for\\ solving the string gravity and supergravity equations\\ in four and higher dimensions}
\author{G.A.~Alekseev}
     \email{G.A.Alekseev@mi.ras.ru}
\affiliation{\centerline{
\hbox{Steklov Mathematical
Institute of the Russian Academy of Sciences,}}\\
\centerline{\hbox{Gubkina str. 8, 119991, Moscow, Russia}}
}

\begin{abstract}
\noindent
The monodromy transform and corresponding integral equation method described here give rise to a general systematic approach for solving integrable reductions of field equations for gravity coupled bosonic dynamics in string gravity and supergravity as well as for pure vacuum gravity in four and higher dimensions. For physically different types of fields in space-times of $D\ge 4$ dimensions with $d=D-2$ commuting isometries -- stationary fields with spatial symmetries, interacting waves or evolution of partially inhomogeneous cosmological models, the string gravity equations govern the dynamics of interacting gravitational, dilaton, antisymmetric tensor and any number $n\ge 0$ of Abelian vector gauge fields (all depending only on two coordinates). The equivalent spectral problem constructed earlier allows to parameterize the entire infinite-dimensional space of (normalized) local solutions of these equations by two pairs of \cal{arbitrary} coordinate-independent holomorphic $d\times d$- and $d\times n$- matrix functions $\{\mathbf{u}_\pm(w),\, \mathbf{v}_\pm(w)\}$  of a spectral parameter $w$ which constitute a complete set of monodromy data for normalized fundamental solution of this spectral problem. The ``direct'' and ``inverse'' problems of such monodromy transform --- calculating the monodromy data for any local solution and constructing the field configurations for any chosen monodromy data  always admit unique solutions. We construct the linear singular integral equations which solve this inverse problem. For any \emph{rational} and \emph{analytically matched} (i.e. $\mathbf{u}_+(w)\equiv\mathbf{u}_-(w)$ and $\mathbf{v}_+(w)\equiv\mathbf{v}_-(w)$) monodromy data the solution of these integral equations and corresponding solution for string gravity equations can be found explicitly. Simple reductions of the space of monodromy data leads to the similar constructions for solving of other integrable symmetry reduced gravity models, e.g. $5D$ minimal supergravity or vacuum gravity in $D\ge 4$ dimensions.
\end{abstract}

\pacs{04.20.Jb, 04.50.-h, 04.65.+e, 05.45.Yv}

\maketitle

\section*{Introduction}
Motivated by various physical reasons, numerous modern studies of strong gravitational fields in different gravity models -- pure gravity as well as the bosonic sectors of string gravity and supergravity in four and higher dimensions,  brought us many interesting discoveries concerning the existence and  properties of a large variety of higher-dimensional space-time structures -- black holes, black rings, black lens, etc. (see, for example, \cite{Emparan-Reall:2008}), which distinguish these models in higher dimensions from those in 4D space-times. It can be expected that the same may concern also physically different types of fields, e.g. interacting waves and cosmological solutions. In all cases, an explicit construction of appropriate solutions can play even more crucial role than it was in four dimensions, where our physical intuition is better adapted to.

The construction of solutions for different gravity models in four and higher dimensions needs a generalization of various methods developed earlier for 4D case to make these available
for solving the field equations in higher dimensions and to include various non-gravitational fields which play important roles in these theories. For these purposes, in particular, it is
very important to find the cases in which the symmetry reduced dynamical equations occur to be completely integrable. Such integrable reductions may arise if space-times of $D$ dimensions admit $D-2$ commuting isometries and therefore, all field components and potentials depend only on two space-time coordinates. Now we have a large experience in the development and application of various solution generating methods for Einstein's field equations in $4 D$ space-times with two commuting isometries (see, e.g., the survey \cite{Alekseev:2010} and the references therein). In $4D$ case, the already developed methods allow, in particular, to construct explicitly the hierarchies of solutions with arbitrary large number of free parameters. These are the $N$-soliton solutions generated on (or, in another words, interacting with) arbitrary chosen backgrounds such as  Belinski and Zakharov vacuum solitons \cite{Belinski-Zakharov:1978} and Einstein - Maxwell solitons \cite{Alekseev:1980}, as well as some classes of non-soliton solutions \cite{Alekseev:1983,
Alekseev:1992, Alekseev-Griffiths:2000}. Besides that, for these equations one can solve the (effectively - two-dimensional) boundary value problems for stationary fields (see \cite{Meinel-Ansorg-Kleinw¨achter-Neugebauer-Petroff:2008}
and the references there) and characteristic initial value problems
\cite{Alekseev:2001, Alekseev-Griffiths:2001, Alekseev-Griffiths:2004}. However, a generalization of these methods for higher dimensional gravity models is not trivial and it may need to overcome some specific difficulties.

In the last two decades, many authors used the Belinski and Zakharov soliton generating transformations for constructing asymptotically flat stationary axisymmetric solutions for pure vacuum 5D gravity. This ``dressing'' method for generating solitons on arbitrary (vacuum) background can be generalized directly from $4D$ to higher
dimensions without any essential changes of the procedure. However for appropriate choice of soliton parameters and practical constructing of particular solutions with desired propeties for the fields of rotating objects in $5D$ space-times, the analysis of the so called ``rod structure'' of soliton solutions suggested in \cite{Emparan-Reall:2002, Harmark:2004} was found very helpful. (Note that many decades ago, the similar diagrams were used for classification of static vacuum solitons in $4D$ case \cite{Alekseev-Belinski:1980}.)

A series of attempts was made also to apply the Belinski and Zakharov construction of the spectral problem to other gravity models in $4D$ as well as to bosonic sector of 5D minimal supergravity. The main problem which arose in these applications and which was not solved there is that the constructions of the Lax pairs (like the Belinski and Zakharov one) had not been supplied by appropriate conditions imposed on the fundamental solutions of the linear systems which would provide the equivalence of the entire spectral problems to the dynamical equations. In these cases, the application of Belinski and Zakharov soliton generating transformations was not successful because the generating soliton solutions left the corresponding coset space (see the comments in \cite{Figueras-Jamsin-Rocha-Virmani:2010}).

On the other hand, the spectral problem constructed in \cite{Alekseev:2009} is \emph{equivalent} to the dynamical equations which govern the bosonic sector of heterotic string effective theory in the space-time of $D$ dimensions with $d=D-2$ commuting isometries.
(These equations include as dynamical variables the metric, dilaton field, antisymmetric potential for the 3-form gauge field and any number $n$ of Abelian vector gauge fields, all depending only on two space-time coordinates.) In $5D$ case with $n=1$, certain constraints imposed on the potentials reduce this system and corresponding equivalent spectral problem to those for the bosonic sector of $5D$ minimal supergravity.

As it was mentioned in \cite{Alekseev:2009}, the spectral problem constructed there possess the monodromy preserving properties, which provide the base for generalization for this case of the monodromy transform approach suggested by the author in \cite{Alekseev:1985} for solving the symmetry reduced vacuum Eisntein equations, electrovacuum Eistein - Maxwell and Einstein - Maxwell - Weyl field equations in General Relativity (see also \cite{Alekseev:1988, Alekseev:2005b}).

  In this paper we present some sketch of the monodromy transform construction for heterotic string gravity model in $D$ dimensions and $5D$ minimal supergravity including
\begin{itemize}
\item{the solution of a ``direct'' problem  -- a parametrization of the whole space of local solutions of symmetry reduced bosonic equations for heterotic string gravity by  pairs of matrix valued coordinate-independent holomorphic functions of spectral parameter $\{\mathbf{u}_\pm^{d\times d}(w),\,\mathbf{v}_\pm^{d\times n}(w)\}$ which arise as the monodromy data of the normalized fundamental solution of the associated spectral problem and which characterize uniquely any local solution;}
\item{the solution of the ``inverse'' problem -- a construction of the local solutions for any given monodromy data. For solution of this problem we construct a system of linear singular integral equations which solution for any given monodromy data always exists and is unique, and it allows to calculate all field components in quadratures;}
\item{explicit construction of infinite hierarchies of multiparametric families of exact solutions for a special class of ``analytically matched'' monodromy data such that $\mathbf{u}_+(w)=\mathbf{u}_-(w)\equiv\mathbf{u}(w)$ and $\mathbf{v}_+(w)=\mathbf{v}_-(w)\equiv \mathbf{v}(w)$. These conditions are equivalent to regularity of the local solution at the boundary which consists of stationary points of the isometry group (the ``axis'' of symmetry). The monodromy data functions $\mathbf{u}(w)$ and $\mathbf{v}(w)$ for this class are closely related to the values of the matrix Ernst potentials on the ``axis''. If $\mathbf{u}(w)$ and $\mathbf{v}(w)$ are chosen as any \emph{rational} functions of $w$, the corresponding solution can be found explicitly.}
\end{itemize}
Finally we note, that the monodromy data play the role of ``coordinates'' in the infinite-dimensional space of local solutions of symmetry reduced dynamical equations. If we impose certain restrictions on the choice of the monodromy data, the integral equation method developed here for solving the for bosonic field equations in heterotic string theory can be used also for constructing solutions for pure vacuum gravity in $D$ dimensions and $5D$ minimal supergravity (the details are expected to be the subject of some later publications).

\section*{Massless bosonic dynamics in string gravity}
The massless bosonic part of string effective action in space-times with $D\ge 4$ dimensions in string frame is
\begin{equation}
\label{StringFrame}
\begin{array}{l}
{\cal S}=\!\!\displaystyle\int e^{-\widehat{\Phi}}\left\{\widehat{R}{}^{(D)}+\nabla_M \widehat{\Phi} \nabla^M\widehat{\Phi}
-\displaystyle\frac 1{12} H_{MNP} H^{MNP}\right.\\[1ex]
\phantom{{\cal S}^{(D)}=\int e^{-\widehat{\widehat{\Phi}}}}\left.
-\dfrac 12\sum\limits_{\mathfrak{p}=1}^n\displaystyle F_{MN}{}^{(\mathfrak{p})} F^{MN\,(\mathfrak{p})}\right\}\sqrt{- \widehat{G}}\,d^{D}x
\end{array}
\end{equation}
where $M,N,\ldots=1,2,\ldots,D$ and $\mathfrak{p}=1,\ldots n$; $\widehat{G}{}_{MN}$ possesses the ``most positive'' Lorentz signature. Metric $\widehat{G}{}_{MN}$ and dilaton field $\widehat{\Phi}$ are related to the metric  $G_{MN}$ and dilaton  $\Phi$ in the Einstein frame as
\begin{equation}
\label{EinsteinFrame}
\widehat{G}{}_{MN}=e^{2\Phi} G_{MN},\qquad
\widehat{\Phi}=(D-2)\Phi.
\end{equation}
The components of a three-form $H$ and two-forms $F{}^{(\mathfrak{p})}$ are determined in terms of antisymmetric tensor field $B_{MN}$ and Abelian gauge field potentials $A_M{}^{(\mathfrak{p})}$ as
\[\begin{array}{l}
H_{MNP}=3\bigl(\partial_{[M} B{}_{NP]}-\sum\limits_{\mathfrak{p}=1}^n A{}_{[M}{}^{(\mathfrak{p})} F_{NP]}{}^{(\mathfrak{p})}\bigr),\\[0.5ex]
F_{MN}{}^{(\mathfrak{p})}=2\,\partial_{[M} A{}_{N]}{}^{(\mathfrak{p})},\qquad B_{MN}=-B_{NM}.
\end{array}
\]

\section*{Space-time symmetry ansatz}
We consider the space-times with $D\ge 4$ dimensions which admit $d=D-2$ commuting Killing vector fields. All field components and potentials are assumed to be functions of only two coordinates  $x^1$ and $x^2$, one of which can be time-like or both are space-like coordinates. We assume also the following structure of metric components
\begin{equation}\label{Metric}
G_{MN}=\begin{pmatrix}g_{\mu\nu}&0\\
0 & G_{ab}
\end{pmatrix}\qquad
\begin{array}{l}
\mu,\nu,\ldots=1,2\\
a,b,\ldots=3,4,\ldots D
\end{array}
\end{equation}
while the components of field potentials take the forms
\begin{equation}\label{BAfields}
B_{MN}=\begin{pmatrix} 0& 0\\
0 & B_{ab}
\end{pmatrix},\qquad
A_M{}^{(\mathfrak{p})}=\begin{pmatrix}0\\ A_a{}^{(\mathfrak{p})}
\end{pmatrix}.
\end{equation}
We choose $x^1,x^2$ so that $g_{\mu\nu}$ takes a conformally flat form
\[g_{\mu\nu}=f \eta_{\mu\nu},\qquad\eta_{\mu\nu}=
\left(\begin{array}{ll}\epsilon_1 & 0\\
0&\epsilon_2\end{array}\right),\qquad\begin{array}{l}\epsilon_1=\pm1\\
\epsilon_2=\pm1\end{array}\]
where $f(x^\mu)>0$ and  the sign symbols $\epsilon_1$ and $\epsilon_2$ allow to consider various types of fields. The field equations imply that the function $\alpha(x^1,x^2)>0$ is  ``harmonic'' one:
\[
\det\Vert G_{ab}\Vert\equiv \epsilon\alpha^2, \qquad \eta^{\mu\nu}\partial_\mu\partial_\nu\alpha=0,\qquad\epsilon=-\epsilon_1 \epsilon_2.
\]
where $\eta^{\mu\nu}$ is inverse to $\eta_{\mu\nu}$, and
therefore, the function $\beta(x^\mu)$ can be defined as ``harmonically'' conjugated  to $\alpha$:
\[\partial_\mu\beta=\epsilon\varepsilon_\mu{}^\nu\partial_\nu\alpha,
\qquad
\varepsilon_\mu{}^\nu=\eta_{\mu\gamma}\varepsilon^{\gamma\nu},\qquad
\varepsilon^{\mu\nu}=\begin{pmatrix}0&1\\-1&0\end{pmatrix}.
\]
Using the functions $(\alpha,\beta)$, we construct a pair $(\xi,\eta)$ of real null coordinates in the hyperbolic case or complex conjugated to each other coordinates in the elliptic case:
\[\left\{\begin{array}{l} \xi=\beta+j\alpha,\\[1ex]
\eta=\beta-j\alpha,\end{array}\right.\quad
j=\left\{\begin{array}{llll}
1,&\epsilon=1&-&\hbox{hyperbolic case},\\[1ex]
i,&\epsilon=-1&-&\hbox{elliptic case.}
\end{array}\right.
\]
In particular, for stationary axisymmetric fields $\xi=z+i\rho$,
$\eta=z-i\rho$, whereas for plane waves or for cosmological solutions $\xi=-x+t$, $\eta=-x-t$, or these may have more complicate expressions in terms of $x^1$, $x^2$.

\section*{Dynamical equations}
The symmetry reduced dynamical equations for the action (\ref{StringFrame}) can be presented in the form of real matrix Ernst-like equations for the string frame matrix variables -- a symmetric $d\times d$-matrix $\mathcal{G}$, antisymmetric $d\times d$-matrix $\mathcal{B}$, a rectangular $d\times n$-matrix $\mathcal{A}$, the scalars $\widehat{\Phi}$ and $\alpha$:
\[\mathcal{G}=e^{2\Phi}\Vert G_{ab}\Vert,\quad \mathcal{B}=\Vert B_{ab}\Vert,\quad \mathcal{A}=\Vert A_a{}^{(\mathfrak{p})}\Vert
\]
which should satisfy the system of equations
\begin{equation}\label{Ernst_equations}
\left\{\begin{array}{l}
\eta^{\mu\nu}\partial_\mu
(\alpha \partial_\nu{\cal E})- \alpha\,\eta^{\mu\nu}(\partial_\mu{\cal E}-2\partial_\mu {\cal A}{\cal A}^T)\mathcal{G}^{-1}\partial_\nu{\cal E}=0,
\\[1ex]
\eta^{\mu\nu}\partial_\mu(\alpha \partial_\nu{\cal A})-\alpha\,\eta^{\mu\nu}
(\partial_\mu{\cal E}-2\partial_\mu {\cal A}{\cal A}^T)\mathcal{G}^{-1}\partial_\nu{\cal A}=0,\\[1ex]
\eta^{\mu\nu}\partial_\mu\partial_\nu\alpha=0,
\end{array}\right.
\end{equation}
where $T$ means a matrix transposition and
\begin{equation}\label{Ernst_potential}
{\cal E}=\mathcal{G}+\mathcal{B}+{\cal A} {\cal A}^T,
\qquad
\det \mathcal{G}=\epsilon\alpha^2e^{2\widehat{\Phi}}.
\end{equation}
The equations (\ref{Ernst_equations}) imply the existence of antisymmetric $d\times d$ potential $\widetilde{\mathcal{B}}$ and $d\times n$ potential $\widetilde{\mathcal{A}}$ defined as
\begin{equation}\label{duals}
\begin{array}{l}
\partial_\mu\widetilde{\mathcal{B}}= -\epsilon\alpha\varepsilon_\mu{}^\nu \mathcal{G}^{-1}(\partial_\nu\mathcal{B}-\partial_\nu{\cal A} {\cal A}^T+{\cal A} \partial_\nu{\cal A}^T)\mathcal{G}^{-1},\\[1ex]
\partial_\mu\widetilde{\mathcal{A}}=-\epsilon\alpha \varepsilon_\mu{}^\nu \mathcal{G}^{-1}\partial_\nu \mathcal{A}+\widetilde{\mathcal{B}}\,\partial_\mu \mathcal{A}.
\end{array}
\end{equation}
The remaining (non-dynamical) part of field equations determines the conformal factor $f$ in quadratures , provided the solution of dynamical equations is found \cite{Alekseev:2009}.

\section*{Equivalent spectral problem}
As it was described in \cite{Alekseev:2009}, the dynamical equations (\ref{Ernst_equations}) admit an \textsl{equivalent} reformulation in terms of the spectral problem for the four $(2 d+n)\times (2 d+n)$-matrix functions depending on two real (in a hyperbolic case) or two complex conjugated (in the elliptic case) coordinates $\xi$ and $\eta$ and a free complex ("spectral") parameter $w\in \mathbb{C}$
\begin{equation}\label{PsiUVW}
\mathbf{\Psi}(\xi,\eta,w),\,\,\mathbf{U}(\xi,\eta),\,\,
\mathbf{V}(\xi,\eta),\,\,\mathbf{W}(\xi,\eta,w)
\end{equation}
which should satisfy the following linear system for $\mathbf{\Psi}$ with algebraic constraints on its matrix coefficients
\begin{equation}\label{LinSys}
\left\{\begin{array}{l}
2(w-\xi)\partial_\xi \mathbf{\Psi}=\mathbf{U}(\xi,\eta) \mathbf{\Psi}\\[2ex]
2(w-\eta)\partial_\eta \mathbf{\Psi}=\mathbf{V}(\xi,\eta) \mathbf{\Psi}
\end{array}\hskip1ex\right\Vert
\hskip1ex\begin{array}{l}
{\bf U}\cdot{\bf U} ={\bf U},\hskip1ex \text{tr}{\bf U}=d \\[2ex]
{\bf V}\cdot{\bf V} ={\bf V},\hskip1ex\text{tr}{\bf V}=d
\end{array}
\end{equation}
It is necessary also that the system (\ref{LinSys}) will possess a symmetric matrix integral ${\mathbf W}_o(w)$ such that
\begin{equation}\label{WCondition}
\left\{\begin{array}{l}
{\bf \Psi}^T {\bf W} {\bf \Psi}={\bf W}_o(w)\\[2ex]
{\bf W}_o^T(w)={\bf W}_o(w)
\end{array}
\hskip1ex\right\Vert\quad
\dfrac{\partial\mathbf{W}}{\partial w}=\mathbf{\Omega},\hskip1ex
\mathbf{\Omega}=\begin{pmatrix}
0&I_d&0\\I_d&0& 0\\0&0&0
\end{pmatrix}
\end{equation}
where $I_d$ is $d\times d$ unit matrix and $\mathbf{\Omega}$ is $(2 d+n)\times (2 d+n)$-matrix. We should impose also the reality conditions
\begin{equation}\label{reality}
\overline{\mathbf{\Psi}(\xi,\eta,\overline{w})}= \mathbf{\Psi}(\xi,\eta,w),\hskip0.5ex
\overline{\mathbf{W}_o(\overline{w})}= \mathbf{W}_o(w),\hskip0.5ex
\mathbf{W}_{(3)(3)}=I_n
\end{equation}
where $\mathbf{W}_{(3)(3)}$  is the lower right $n\times n$-block of $\mathbf{W}$ which, in accordance with  (\ref{PsiUVW})--(\ref{WCondition}), is a constant matrix and therefore the last condition in (\ref{reality}) is pure gauge.

\section*{Field variables and potentials}
As it can be shown by direct calculations (the detail will be published elsewhere), the conditions (\ref{PsiUVW})--(\ref{WCondition}) imply, in particular, that $\mathbf{W}$ possess a special structure
\begin{equation}\label{GW}
\begin{array}{l}
\mathbf{W}=(w-\beta)\mathbf{\Omega}+\mathbf{G},\qquad\text{where}\\[1ex]
\mathbf{G}=\begin{pmatrix}
\epsilon\alpha^2\mathcal{G}^{-1}-\widetilde{\mathcal{B}} \mathcal{G}\widetilde{\mathcal{B}} +\widetilde{\mathcal{A}}\widetilde{\mathcal{A}}{}^T
&\widetilde{\mathcal{B}}\mathcal{G}+ \widetilde{\mathcal{A}}\mathcal{A}{}^T
&\widetilde{\mathcal{A}}\\[1ex]
-\mathcal{G} \widetilde{\mathcal{B}}
+{\mathcal{A}}\widetilde{\mathcal{A}}{}^T
&\mathcal{G}+\mathcal{A}\mathcal{A}{}^T& \mathcal{A}\\[1ex]
\widetilde{\mathcal{A}}{}^T&\mathcal{A}^T&I_n
\end{pmatrix}
\end{array}
\end{equation}
and $\alpha=(\xi-\eta)/2 j$, $\beta=(\xi+\eta)/2$; $d\times d$-matrix blocks $\mathcal{G}$ and $\widetilde{\mathcal{B}}$
are symmetric and antisymmetric respectively and, together with $d\times n$-matrices $\mathcal{A}$ and $\widetilde{\mathcal{A}}$, these satisfy (\ref{Ernst_equations}) -- (\ref{duals}). (The corresponding expressions for
$\mathbf{U}$ and $\mathbf{V}$ also have correct forms described in \cite{Alekseev:2009}.) This allows to calculate all field components and potentials for any solution (\ref{PsiUVW}) of our spectral problem.

\section*{Global symmetries}
The spectral problem (\ref{PsiUVW})--(\ref{reality}) admits global symmetry transformations (including the discrete ones)
\begin{equation}\label{Symmetries}
\left.\begin{array}{l}
\mathbf{U}\to\mathbf{A}\,\mathbf{U}\,\mathbf{A}^{-1},\quad
\mathbf{V}\to\mathbf{A}\,\mathbf{V}\,\mathbf{A}^{-1}\\[1.5ex]
\mathbf{\Psi}\to \mathbf{A}\mathbf{\Psi},\quad
\mathbf{W}\to(\mathbf{A}^T){}^{-1}\mathbf{W}\,\mathbf{A}^{-1}
\end{array}\hskip1ex\right\Vert\hskip1ex
\begin{array}{l}
\mathbf{A}^T\mathbf{\Omega}\mathbf{A}=\mathbf{\Omega}\\[1.5ex]
\mathbf{W}_{(3)(3)}\equiv I_n
\end{array}
\end{equation}
where the real constant matrix $\mathbf{A}$ is determined by two invariance conditions shown just above on the right. Some of these symmetries are not pure gauge and generate physically different solutions from a given one. We use further two subgroups of pure gauge transformations:
\begin{equation}
\mathbf{A}^{-1}=\begin{pmatrix}I_d&0&0\\
-\omega&I_d&0\\
\widetilde{a}{}^T & a^T & I_n \end{pmatrix}
\hskip1ex \text{and}\hskip1ex
\mathbf{A}^{-1}=\begin{pmatrix}(L_d^T)^{-1}&0&0\\
0&L_d&0\\
0&0& S_n\end{pmatrix}
\end{equation}
where $\omega$ is antisymmetric $d\times d$-matrix, $a$ and $\widetilde{a}$ are $d\times n$-matrices, $L_d\in GL(d,\mathbb{R})$, $S_n\in O(d)$. These generate the following changes of the field variables and potentials:
\begin{equation}
\begin{array}{l}
\mathcal{G}^\prime=\mathcal{G}\\ \widetilde{\mathcal{B}}{}^\prime=\widetilde{\mathcal{B}}+\omega\\
\mathcal{A}^\prime=\mathcal{A}+a\\
\widetilde{\mathcal{A}}{}^\prime=\widetilde{\mathcal{A}}+\omega \mathcal{A}+\widetilde{a}
\end{array}\quad\text{and}\qquad
\begin{array}{l}
\mathcal{G}^\prime=L_d^T \mathcal{G} L_d,\\ \widetilde{\mathcal{B}}{}^\prime=L_d^{-1}\widetilde{\mathcal{B}} (L_d^{-1})^T\\
\mathcal{A}^\prime=L_d^T\mathcal{A} S_n \\
\widetilde{\mathcal{A}}{}^\prime=L_d^{-1}\widetilde{\mathcal{A}} S_n
\end{array}
\end{equation}
Another pure gauge transformation changes only the choice of the fundamental solution $\mathbf{\Psi}(\xi,\eta,w)$ of (\ref{LinSys}) using arbitrary nondegenerate matrix $\mathbf{C}(w)$
\begin{equation}\label{PsiTransform}
\mathbf{\Psi}(\xi,\eta,w)\to \mathbf{\Psi}(\xi,\eta,w)\mathbf{C}(w)
\end{equation}
We need also a set of non-gauge discrete transformations which interchange in (\ref{Symmetries}) the $k$th with $(k+d)$th rows and  $k$th with $(k+d)$th columns of a matrix ($1\le k\le d$):
\begin{equation}\label{Zsymmetries}
\mathbf{A}=\mathbf{Z}_{(k,d+k)},\qquad  \mathbf{Z}_{(k,d+k)}=\mathbf{Z}^T_{(k,d+k)}=\mathbf{Z}^{-1}_{(k,d+k)}
\end{equation}

\section*{The space of normalized local solutions}
The main object of our present consideration is the space of all (normalized) local solutions of (\ref{Ernst_equations}), i.e. all solutions which components and potentials are holomorphic (as functions of two complex variables $\xi$ and $\eta$) in some  local domains near a chosen "initial" point $(\xi_o,\eta_o)$. Therefore, we have to consider a general solution of  (\ref{PsiUVW})--(\ref{reality}) locally in $\xi$ and $\eta$, but on the entire spectral plane:
\[(\xi,\eta)\in (\Omega_{\xi_o}\times\Omega_{\eta_o}),\qquad w\in\overline{\mathbb C}
\]
where $\Omega_{\xi_o}$, $\Omega_{\eta_o}$ are some local regions on the complex plane near $\xi_o$ and $\eta_o$ respectively such that the matrices $\mathbf{U}$ and $\mathbf{V}$ are holomorphic functions of $(\xi,\eta)$ in $\Omega_{\xi_o}\times\Omega_{\eta_o}$.

Besides that, without any loss of generality we impose on the field components and auxiliary matrix functions a set of normalization conditions which provide unambiguous correspondence between local solutions of (\ref{Ernst_equations}) and the solutions of our spectral problem. For this we set
\begin{equation}\label{Normalization}
\begin{array}{l}
\mathbf{\Psi}(\xi_o,\eta_o,w)=\mathbf{I},\qquad \mathbf{W}_o(w)=(w-\beta_o)\mathbf{\Omega}+\mathbf{G}_o\\[2ex]
\mathbf{G}_o=\begin{pmatrix}
\epsilon\alpha_o^2\mathcal{G}_o^{-1}&0&0\\[1ex]
0&\mathcal{G}_o&0\\[1ex]
0&0&I_n
\end{pmatrix},\qquad
\begin{array}{l}
\mathcal{G}(\xi_o,\eta_o)=\mathcal{G}_o,\\
\mathcal{B}(\xi_o,\eta_o)=0,\\
\mathcal{A}(\xi_o,\eta_o)=0,\\
\widetilde{\mathcal{A}}(\xi_o,\eta_o)=0
\end{array}
\end{array}
\end{equation}
where $\alpha_o=(\xi_o-\eta_o)/2 j$, $\beta_o=(\xi_o+\eta_o)/2$ and $d\times d$-matrix
\[\mathcal{G}_o=\left\{\begin{array}{lcll}
\text{diag} \{\epsilon,1,\,\ldots,\,1\}&-& \text{for}& \alpha_o\ne 0\\
\text{diag} \{\epsilon,1,\ldots,\pm 1\}&-& \text{for}& \alpha_o=0
\end{array}\right.
\]
All these normalization conditions can be achieved using the transformations (\ref{Symmetries})--(\ref{PsiTransform}). Note, however, that in the case $\alpha_o=0$ (as, e.g. in stationary axisymmetric case with the initial point chosen on the axis), the finite expression for $\mathbf{G}_o$ arises from the case $\alpha_o\ne 0$ as the limit for $\alpha_o\to 0$. To get the non-degenerate $\mathcal{G}_o$ in this case, one needs to use the transformation (\ref{Symmetries}), (\ref{Zsymmetries}) which may lead to a change of signature of $\mathcal{G}_o$, and after a construction of the solution, we have to perform the inverse transformation.

\section*{General analytic structure of $\mathbf{\Psi}$ on $w$-plane.}
\paragraph*{Global structure of $\mathbf{\Psi}$ on the spectral plane.} As it is easy to see from the structure of (\ref{LinSys}) and (\ref{Normalization}), the normalized fundamental solution $\mathbf{\Psi}(\xi,\eta,w)$ of the linear system (\ref{LinSys}) possess in general only four singular points $w=\xi_o$, $w=\xi$, $w=\eta_o$, $w=\eta$ which are the branching points of the orders $(\frac12,-\frac12,\frac12,-\frac12)$ respectively for $\mathbf{\Psi}(\xi,\eta,w)$ and opposite for its inverse. To select a holomorphic branch of $\mathbf{\Psi}(\xi,\eta,w)$, we make on the $w$-plane two \underline{local} cuts, one of which $L_+$ goes from $w=\xi_o$ to $w=\xi$ and another one $L_-$ from $w=\eta_o$ to $w=\eta$. We assume that each of these cuts belongs to the corresponding local region $\Omega_+=\{w\,\vert\,w=\xi\in \Omega_{\xi_o}\}$ and $\Omega_-=\{w\,\vert\,w=\eta\in \Omega_{\eta_o}\}$.
\vskip-2ex
\[\begin{matrix}
\hskip10ex
\Omega_{\scriptscriptstyle-} \hskip20ex \Omega_{\scriptscriptstyle+} \\[0ex]
\hskip1ex
L_{\scriptscriptstyle-} \hskip20ex L_{\scriptscriptstyle+} \\[-1.75ex]
\ellipse{65}{30}\hskip23ex\ellipse{65}{30}\\[-3ex]
{\vrule width10ex height0.05ex depth0.05ex}
{\vrule width0.05ex height0.5ex depth0.5ex}
{\vrule width10ex height0.15ex depth0.15ex}
{\vrule width0.05ex height0.5ex depth0.5ex}
{\vrule width13ex height0.05ex depth0.05ex}
{\vrule width0.05ex height0.5ex depth0.5ex}
{\vrule width10ex height0.15ex depth0.15ex}
{\vrule width0.05ex height0.5ex depth0.5ex}
{\vrule width10ex height0.05ex depth0.05ex}\\
\noalign{\vskip-0.3ex}
\hskip1ex \eta_o
\hskip8ex \eta \hskip12ex \xi \hskip9ex \xi_o
\end{matrix}
\]
\vskip-3ex
\begin{description}
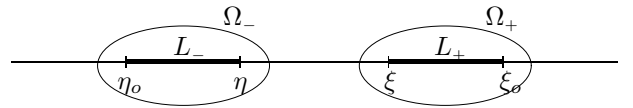
\item[{\rm Fig.1}]{\small Cuts on the $w$-plane in hyperbolic ($\epsilon=1$) case. In the elliptic ($\epsilon=-1$) case, the cuts are located symmetrically to each other with respect to real axis of $w$-plane.}
\end{description}

\paragraph*{Local structure of $\mathbf{\Psi}$ near the cuts.} It can be shown (see \cite{Alekseev:2005b}) that the normalized matrix function $\mathbf{\Psi}$ and its inverse possess near the cuts $L_\pm$ the following structures
\begin{equation}
\begin{array}{l}\label{LocalStructure}
{\bf \Psi}=\left\{\!\begin{array}{ll}
\lambda_+^{-1}\,\mathbf{\bpsi}_+(\xi,\eta,w)\cdot\,
{\bf k}_+(w)+{\bf M}_+(\xi,\eta,w),&w\in\Omega_+\\[0.5ex]
\lambda_-^{-1}\,{\bpsi}_-(\xi,\eta,w)\cdot\,
{\bf k}_-(w)+{\bf M}_-(\xi,\eta,w),&w\in\Omega_-
\end{array}\right.\\[3ex]
{\bf \Psi}^{-1}\! =\!\!\left\{\!\begin{array}{lcl}
\lambda_+\,\,{\bf l}_+(w)\,\cdot\,{\bphi}_+(\xi,\eta,w) +{\bf N}_+(\xi,\eta,w),
&w\in\Omega_+\\[0.5ex]
\lambda_-\,\,{\bf l}_-(w)\,\cdot\,{\bphi}_-(\xi,\eta,w) +{\bf N}_-(\xi,\eta,w),
&w\in\Omega_-\end{array}\right.
\end{array}
\end{equation}
where ``$\cdot$'' means mutrix multiplication and the functions $\lambda_\pm$ are defined by the expressions:
$\lambda_{\scriptscriptstyle\pm}(w=\infty)=1$ and
\[
\lambda_+ = \sqrt{(w-\xi)/(w-\xi_{\scriptscriptstyle{0}})},\quad \lambda_- = \sqrt{(w-\eta)/
(w-\eta_{\scriptscriptstyle{0}})};
\]
all other ``fragments'' of the local structures (\ref{LocalStructure}) --- $d\times (2d+n)$-matrix functions $\mathbf{k}_\pm(w)$ and  $\bphi_\pm(\xi,\eta,w)$, \hskip1ex $ (2d+n)\times d$-matrix functions $\mathbf{l}_\pm(w)$ and  $\bpsi_\pm(\xi,\eta,w)$ as well as $(2 d+n)\times (2d+n)$-matrix functions $\mathbf{M}_\pm(\xi,\eta,w)$ and $\mathbf{N}_\pm(\xi,\eta,w)$ are holomorphic in $\Omega_+$ or $\Omega_-$  respectively and should satisfy there the algebraic relations
\begin{equation}
\begin{array}{lcclccl}\label{kNrelations}
\mathbf{k}\cdot\mathbf{N}=0,&&&\mathbf{N}\cdot \bpsi=0,&&&\mathbf{l}(w)=\Sigma_o^2\, \mathbf{W}_o^{-1}(w)\cdot \mathbf{k}^T(w),\\[2ex]
\mathbf{M}\cdot\mathbf{l}=0,&&&\bphi\cdot \mathbf{M}=0,&&&\Sigma_o^2\equiv(w-\xi_o)(w-\eta_o).
\end{array}
\end{equation}
Here and somewhere below, we use instead of a pair of functions with indices ``$+$'' and ``$-$''  defined and holomorphic in $\Omega_+$ and $\Omega_-$ respectively, a one function defined in  $\Omega_+\cup\Omega_-$ and represented in $\Omega_+$ and $\Omega_-$  by the corresponding functions with the indices ``$+$'' or ``$-$''.

\section*{Monodromy data}

The coordinate-independent functions $\mathbf{k}_\pm(w)$ and $\mathbf{l}_\pm(w)$ play a very important role in our construction. These functions determine the monodromy of $\mathbf{\Psi}$ on the cuts $L_\pm$ because the analytical continuation of $\mathbf{\Psi}$ along a  simple path which goes from one edge of a cut $L_+$ or $L_-$ to its another edge, leads to the linear transformations of $\mathbf{\Psi}\,\to\, \mathbf{\Psi}\cdot\mathbf{T}_\pm(w)$ with the matrices $\mathbf{T}_\pm(w)$ which structure follows immediately from (\ref{LocalStructure}):
\vskip-3ex
\begin{equation}\label{Tmatrices}\mathbf{T}_\pm(w)=\mathbf{I}-2\,\, \mathbf{l}_\pm(w)\cdot\bigl(\mathbf{k}_\pm(w)\cdot\mathbf{l}_\pm(w)\bigr)^{-1} \!\!\cdot\mathbf{k}_\pm(w)
\end{equation}
where one should choose only upper or only lower signs. Note also that $\mathbf{T}^2_\pm(w)\equiv \mathbf{I}$.

In what follows, we are going to characterize unambiguously the solutions of (\ref{Ernst_equations}) by a complete set of independent functional parameters in (\ref{Tmatrices}) which we call the monodromy data of a solution. For this we note that in view of (\ref{kNrelations}), $\mathbf{l}_\pm(w)$ can be expressed in terms of $\mathbf{k}_\pm(w)$ and therefore, $\mathbf{T}_\pm(w)$ are completely determined by  $\mathbf{k}_\pm(w)$. On the other hand,
it is easy to see, that $\mathbf{\Psi}$ remains unchanged after the transformations
\[\mathbf{k}_\pm(w)\to \mathbf{c}_\pm(w)\cdot\mathbf{k}_\pm(w),\quad \bpsi_\pm(w)\to \bpsi_\pm(w)\cdot\mathbf{c}_\pm^{-1}(w)
\]
where $\mathbf{c}_\pm(w)$ are arbitrary non-degenerate $d\times d$-matrix functions holomorphic in $\Omega_+$ and $\Omega_-$ respectively. This means that not all components of $\mathbf{k}_\pm(w)$ are important in (\ref{Tmatrices}), and to reduce this ambiguity we should consider $\mathbf{k}_\pm(w)$ as functions which take the values in Grassmann manifold $\mathbb{G}_{d,2 d+n}(\mathbb{C})$. The last restriction on $\mathbf{k}_\pm(w)$ arises from the reality condition (\ref{reality}) and it takes the form $\overline{\mathbf{k}_\pm(\overline{w})}=\mathbf{k}_\pm(w)$ in the hyperbolic case ($\epsilon=1$) or $\overline{\mathbf{k}_\pm(\overline{w})}=\mathbf{k}_\mp(w)$ in the elliptic case ($\epsilon=-1$). Just these $\mathbb{G}_{d,2 d+n}$-valued functions $\mathbf{k}_\pm(w)$ satisfying the above reality condition, parametrize unambiguously the whole space of local solutions of (\ref{Ernst_equations}).

To specify the monodromy data as arbitary independent functional parameters, we chose the affine parametrization of the elements of $\mathbb{G}_{d,2 d+n}(\mathbb{C})$. For this, we choose $\mathbf{c}_\pm(w)$, so that any $d$ linear independent columns of $\mathbf{k}_+(w)$ as well as of $\mathbf{k}_-(w)$ will be reduced to a unite matrix $I_d$. If these $d$ linear independent columns
can be chosen to be located at the same positions in $\mathbf{k}_+(w)$ and $\mathbf{k}_-(w)$ (what is the generic situation, though there is some exceptional subspace of solutions for which $\mathbf{k}_\pm(w)$ can not be presented in the form given below, and these solutions should be considered separately), the transformations (\ref{Zsymmetries}) can be used to locate these columns at first $d$ positions. Then $\mathbf{k}_\pm(w)$ take the forms
\begin{equation}\label{upmvpm}
\mathbf{k}_\pm(w)=\{\mathbf{I}_d,\,\,
\mathbf{u}_\pm(w),\,\,
\mathbf{v}_\pm(w)\}
\end{equation}
where two $d\times d$-matrix functions $\mathbf{u}_\pm(w)$ and two $d\times n$-matrix functions $\mathbf{v}_\pm(w)$, which are holomorphic in $\Omega_+$ and $\Omega_-$ respectively and satisfying the  reality conditions, play the role of ``coordinates'' in the infinite-dimensional space of local solutions for (symmetry reduced) bosonic sector of the effective heterotic string theory.

\section*{Linear singular integral equations}
In accordance with the well known theorems of complex analysis, the analytical structure of  $\mathbf{\Psi}(\xi,\eta,w)$ allows to present it and its inverse as the Cauchy-type integrals
\begin{equation}\label{Cauchy}\mathbf{\Psi}=\mathbf{I}+\dfrac 1{i\pi}
\displaystyle\int\limits_L\dfrac {[\mathbf{\Psi}]_\zeta}{\zeta-w}\,d\zeta,\quad
\mathbf{\Psi}^{-1}=\mathbf{I}+\dfrac 1{i\pi}
\displaystyle\int\limits_L\dfrac {[\mathbf{\Psi}^{-1}]_\zeta} {\zeta-w}\,d\zeta
\end{equation}
where $[\mathbf{\Psi}]_\zeta$ and $[\mathbf{\Psi}^{-1}]_\zeta$ denote the jumps (i.e. half of the difference of left and right limits) of $\mathbf{\Psi}$ and $\mathbf{\Psi}^{-1}$ at the points $\zeta\in L$. The ``continuous parts'' $\{\mathbf{\Psi}\}_\zeta$ and $\{\mathbf{\Psi}^{-1}\}_\zeta$  (i.e. half of the sums of left and right limits) of $\mathbf{\Psi}$ and $\mathbf{\Psi}^{-1}$ on $L$ are determined by the integrals of the form (\ref{Cauchy}) in which, however, we put $w=\tau\in L$ and the singular integrals should be considered as the Cauchy principal value integrals (Sokhotski-Plemelj formula) \cite{Gakhov:1977}.
From (\ref{LocalStructure}) we obtain at the point $\zeta\in L=L_+\cup L_-$:
\begin{equation}\label{Jumps}\begin{array}{l}
[\mathbf{\Psi}]_\zeta=[\lambda^{-1}]_\zeta\,\bpsi(\xi,\eta,\zeta)\cdot\mathbf{k}(\zeta),\qquad \{\mathbf{\Psi}\}_\zeta=\mathbf{M}(\xi,\eta,\zeta)\\[1ex]
[\mathbf{\Psi}^{-1}]_\zeta=[\lambda]_\zeta\,\,\mathbf{l}(\zeta)\cdot\bphi(\xi,\eta,\zeta),\qquad \{\mathbf{\Psi}^{-1}\}_\zeta=\mathbf{N}(\xi,\eta,\zeta)
\end{array}
\end{equation}
where we took into account that $\{\lambda_\pm\}\equiv 0$ on $L_\pm$. Using these expressions for $\mathbf{M}$ and $\mathbf{N}$ in terms of Couchy principal value integrals in the relations (\ref{kNrelations}) $\mathbf{k}\cdot\mathbf{N}=0$ and $\mathbf{M}\cdot\mathbf{l}=0$, we obtain two systems
\begin{eqnarray}
-\dfrac1{i\pi}\displaystyle\vpint\limits_L \dfrac{[\lambda]_\zeta\bigl(\mathbf{k}(\tau)\cdot \mathbf{l}(\zeta)\bigr)}{\zeta-\tau}\,\, \cdot \bphi(\xi,\eta,\zeta)\,d\zeta=\mathbf{k}(\tau)\label{phiequation}\\[1ex]
-\dfrac1{i\pi}\displaystyle\vpint\limits_L \bpsi(\xi,\eta,\zeta)\cdot \dfrac{[\lambda^{-1}]_\zeta\bigl(\mathbf{k}(\zeta)\cdot \mathbf{l}(\tau)\bigr)}{\zeta-\tau}\,\, \,d\zeta=\mathbf{l}(\tau)\label{psiequation}
\end{eqnarray}
which are the linear singular integral equations for matrices $\bphi$ and $\bpsi$, provided the monodromy data are given.

The matrix equations (\ref{phiequation}) and (\ref{psiequation}) are equivalent to each other, and for constructing solutions it is enough to solve only one of them. For pure technical reasons we choose (\ref{phiequation}) as the  basic one. The theory of such equations is well developed \cite{Gakhov:1977}, and it allows to show \cite{Alekseev:2005b} that the solution of this integral equation (at least, for $(\xi,\eta)$ close enough to $(\xi_o,\eta_o)$) always exists for any choice of the monodromy data  $\{\mathbf{u}(w),\,\mathbf{v}(w)\}$.

\section*{Calculation of the field components}
For any given monodromy data and the corresponding solution $\bphi(\xi,\eta,w)$ of the linear integral equation (\ref{phiequation}) the field components can be determined in quadratures.
In accordance with (\ref{Cauchy}) and (\ref{Jumps}), for $w\to\infty$ we have
\[\mathbf{\Psi}^{-1}=\mathbf{I}-w^{-1} \mathbf{R}+\ldots,\hskip1ex
\mathbf{R}=\dfrac 1{i\pi}
\displaystyle\int\limits_L [\lambda]_\zeta\,\,\mathbf{l}(\zeta)\cdot\bphi(\xi,\eta,\zeta) \,d\zeta
\]
where $(2d+n)\times (2d+n)$-matrix $\mathbf{R}(\xi,\eta)$
is determined by the integral over $L_+\cup L_-$. Then we obtain from (\ref{PsiUVW})--(\ref{reality})
\[\mathbf{U}=2 \partial_\xi \mathbf{R},\quad \mathbf{V}=2 \partial_\eta \mathbf{R},\quad
\mathbf{W}=\mathbf{W}_o(w)-\mathbf{\Omega}\cdot \mathbf{R}- \mathbf{R}^T\cdot \mathbf{\Omega}
\]
Due to (\ref{GW}), all field components can be calculated algebraically in terms of the components of the matrix $\mathbf{R}$.

\section*{A class of analytically matched monodromy data}
Solutions with analytically matched monodromy data
\begin{equation}\label{RegAx}
\mathbf{u}_+(w)=\mathbf{u}_-(w)\equiv \mathbf{u}(w),\hskip1ex \mathbf{v}_+(w)=\mathbf{v}_-(w)\equiv \mathbf{v}(w)
\end{equation}
possess the following important properties:

(i) the condition (\ref{RegAx}) is equivalent to regularity of the boundary $\alpha=0$ (e.g., of the ``axis of symmetry'');

(ii) if we locate the initial point ($\xi_o,\eta_o$) on some part of the ``axis'' $\alpha=0$, the analytically matched monodromy data can be expressed in terms of the values of the Ernst potentials on this part of the axis (provided $\mathcal{G}_o=\text{diag} \{\epsilon,1,\ldots,\pm 1\}$ - see the comments to (\ref{Normalization})):
\begin{equation}\label{AxData}\mathbf{u}(\beta)=-\displaystyle\frac{{\cal E}(\beta)-{\cal E}(\beta_o)}{2(\beta-\beta_o)},\hskip1ex
\mathbf{v}(\beta)=-\displaystyle\frac{{\cal A}(\beta)-{\cal A}(\beta_o)} {(\beta-\beta_o)}
\end{equation}

(iii) for any choice of $\mathbf{u}(w)$ and $\mathbf{v}(w)$ as \textit{rational} functions of $w$ the corresponding solution of the field equations can be found explicitly using the inversion rule for Cauchy principal value integrals (Poincare-Bertrand formula \cite{Gakhov:1977}) which takes in our context the form
\[\dfrac1{i\pi}\vpint\limits_L\!\!\dfrac{[\lambda]_\zeta}{\zeta-\tau}
A(\zeta)d\zeta=B(\tau)\hskip0.25ex
\Leftrightarrow\hskip0.25ex
A(\tau)=\dfrac1{i\pi}\vpint\limits_L\!\!\dfrac{[\lambda^{-1}]_\zeta}{\zeta-\tau}
B(\zeta)d\zeta
\]
for any function $A(\tau)$ which satisfy the H\"older condition on $L$ and is bounded on its ends, as well as the elementary theory of residues which allow to reduce these integral equations to an algebraic system.

Many physically important stationary axisymmetric solutions (such as, e.g. black holes in 4D case or black holes, black rings, etc. in 5D case) belong to this class and the integral equation method presented above allows to construct their multparametric generalizations.

We note also that the fields with analytically not matched monodromy data ($\mathbf{u}_+(w)\ne\mathbf{u}_-(w)$ and/or $\mathbf{v}_+(w)\ne\mathbf{v}_-(w)$) also include physically important types of fields, e.g. colliding waves or cosmological models which possess the singularities at $\alpha=0$. Methods for constructing such solutions also can be developed \cite{Alekseev-Griffiths:2000}.

\section*{5D Minimal supergravity}
The equations for the bosonic fields in $5D$ minimal supergravity which are determined by the action
\[S^{(5)}=\displaystyle{\int}\left\{R^{(5)}\ast\,{\bf 1}-2\ast F\wedge F+\dfrac 8{3\sqrt{3}}\,F\wedge F\wedge A\right\}\]
coincide with $D=5$ and $n=1$ bosonic equations of heterotic string gravity (\ref{StringFrame}) if we impose there the constraints
$\Phi=0$ and $H_{ABC}=\ast\,F_{ABC}$ with a subsequent rescaling $F_{AB}\to (2/\sqrt{3}) F_{AB}$. In our present context, these  constraints are equivalent to the relations
\begin{equation}\label{SG}
\det \mathcal{G}=\epsilon\alpha^2,\quad
\widetilde{\mathcal{B}}^{ab}=-\epsilon^{abc} \mathcal{A}_c,\quad \mathcal{B}_{ab}=-\epsilon_{abc} \widetilde{\mathcal{A}}^c
\end{equation}
where $\epsilon^{abc}$ is the Levi-Civita symbol on the orbits of isometry group with coordinates $\{x^3,x^4,x^5\}$; $\epsilon^{345}=\epsilon_{345}=1$.
Thus, the space of solutions of $5D$ minimal supergravity is embedded into the space of solutions of heterotic string gravity. Therefore,   to construct the solutions for $5D$ minimal supergravity, we have to impose the appropriate constraints on the choice of monodromy data. This can be done directly for the class of analytically matched data taking into account  (\ref{AxData}), however, the general class of monodromy data needs more complicate analysis.

\section*{Vacuum gravity in D dimensions}
More strong restrictions imposed on the general monodromy data for heterotic string gravity in $D$ dimensions
\begin{equation}\label{Vacuum}
\det \mathcal{G}=\epsilon\alpha^2,\qquad
\mathcal{B}=0,\qquad \mathcal{A}=0
\end{equation}
leads to the space of vacuum solutions in $D$ dimensions.

\section*{Solutions and their monodromy data}
We consider here some very simple examples of $5D$ solutions. The first of them is the Minkowski metric
\begin{equation}\label{Minkowski}
ds^2=-dt^2+d\rho_1^2+d\rho_2^2+\rho_1^2 d\varphi^2+\rho_2^2 d\psi^2
\end{equation}
with $\alpha=\rho_1\rho_2$, $\beta=z=\dfrac12(\rho_2^2
-\rho_1^2)$. With the point of normalization on the ``axis'' $\rho_2=0$ at $\rho_1=\sqrt{-2 z_o}$, the (analytically matched) monodromy data for (\ref{Minkowski}) are:
\[\mathbf{u}=diag\{0,\,-\dfrac1{2 z_o},\,\dfrac1{2 z_o}\},\qquad \mathbf{v}=0\]
The four-parameteric solution for a charged rotating black hole in $5D$ minimal supergravity \cite{Cvetic-Youm:1996} also possess analytically matched monodromy data with one pole: $\mathbf{u}(w)=\mathbf{u}_0+\mathbf{u}_1/(w-h_1)$ and $\mathbf{v}(w)=\mathbf{v}_1/(w-h_1)$ where the matrices $\mathbf{u}_0$ and $\mathbf{u}_1$, the vector $\mathbf{v}_1$  and the scalar $h$ are real and constant. For simplicity, we show their structure for non-rotating case -- $5D$ Reissner-Nordstr\"om solution:
\[\mathbf{u}(w)=
\begin{pmatrix}
\frac{a_0}{w-h}&0&0\\
0&b_0&c_0\\
0&-c_0&d_0
\end{pmatrix},\qquad
\mathbf{v}(w)=
\begin{pmatrix}
\frac{p_0}{w-h}\\
0\\
0
\end{pmatrix}
\]
where the parameters $a_0$, $b_0$, $c_0$, $d_0$, $p_0$ and $h$  depend on the mass $m$ and charge $q(= s\, m)$  parameters of a black hole and the position $z=z_0$ of the point of normalization:
\[\begin{array}{l}
a_0=\dfrac{m[m-2 z_0(1+2 s^2)]}{4 z_0^2-m^2},\quad d_0=\dfrac{(2 z_0-m)^2-8m s^2 z_0} {2(z_0-h)(4z_0^2-m^2)}\\[2ex]
b_0=-\dfrac1{2(z_0-h)},\quad
c_0=\dfrac{p_0}{z_0-h},\quad p_0=\dfrac{m s\sqrt{1+s^2}}{\sqrt{4 z_0^2-m^2}},\\[3ex]
h=m(1+2 s^2)/2
\end{array}
\]

\section*{Concluding remarks}
For a conclusion we note, that the infinite hierarchies of multiparametric families of solutions which can be constructed explicitly using the approach described above for pure vacuum gravity or heterotic string gravity in $D$ dimensions as well as for $5D$ minimal supergravity
include many known solutions and allow to construct their multiparametric generalizations. In particular, this extends considerably the classes of soliton solutions in these gravity models. However, a lot of work remains and is necessary to identify the most physically interesting solutions and to analyse their physical and geometrical properties.

\section*{Acknowledgements}
The author express his deep thanks to the Institut des Hautes Etudes Scientifiques (Bures-sur-Yvette, France) for hospitality during the author's visits when a part of this work was made and to Thibault Damour for numerous stimulating discussions of interrelations between various solution generating techniques for Einstein's field equations. This work was supported in parts by the Russian Foundation for Basic Research (grants
11-01-00034, 11-01-00440) and the program "Fundamental problems of Nonlinear Dynamics" of  Russian Academy of Sciences.


\begin{thebibliography}{99}

\bibitem{Emparan-Reall:2008} R.~Emparan and H.~Reall, Living Rev. Relativity 11, 6 (2008).
\bibitem{Alekseev:2010} G.A.~Alekseev,
Proceedings of the Twelfth Marcel Grossmann Meeting on General Relativity, edited by T.~Damour, R.T.~Jantzen and R.~Ruffini, World Scientific, Singapore, Part A, Plenary and Review talks, p. 645 - 666, (2011) [arXiv:1011.3846v1 [gr-qc]]
\bibitem{Belinski-Zakharov:1978} V.A. Belinski and V.E. Zakharov, \textit{Sov.\ Phys.\ JETP} {\bf 48}, 985 (1978); {\bf 50}, 1 (1979).
\bibitem{Alekseev:1980} G.A. Alekseev, JETP Lett., {\bf 32}, 277 (1980).
\bibitem{Alekseev:1983} G.A. Alekseev, Soviet Phys. Dokl. {\bf 28} 17 -- 19 (1983).
\bibitem{Alekseev:1992}  G. A. Alekseev, in Abstracts of contributed papers 13th International Conference on General Relativity and
Gravitation, eds. P.W. Lamberty and O.E. Ortiz (Huerta Grande, Cordoba, Argentina, 1992), p.3.
\bibitem{Alekseev-Griffiths:2000} G.A.Alekseev, J.B.Griffiths, Phys.Rev.Lett. {\bf 84}, p. 5247-5250, (2000); [arXiv:0004034 [gr-qc]]
\bibitem{Meinel-Ansorg-Kleinw¨achter-Neugebauer-Petroff:2008} Meinel R, Ansorg M, Kleinw¨achter A, Neugebauer G and Petroff D  \textit{Relativistic Figures of
Equilibrium} (Cambridge, UK: Cambridge University Press, 2008).
\bibitem{Alekseev:2001}  G.A.~Alekseev, Theoretical and Mathematical Physics, {\bf 129}, 1466-1483 (2001).
\bibitem{Alekseev-Griffiths:2001} G.A.~Alekseev and J.B.~Griffiths, Phys. Rev. Lett. {\bf 87} 221101 (2001);
\bibitem{Alekseev-Griffiths:2004} G.A.~Alekseev and J.B.~Griffiths, Class. Quantum Grav. {\bf 21} 5623–5654 (2004).
\bibitem{Emparan-Reall:2002} R. Emparan and H. S. Reall, Phys. Rev. D 65, 084025 (2002).
\bibitem{Harmark:2004} T.~Harmark, Phys. Rev. D 70, 124002 (2004).
\bibitem{Alekseev-Belinski:1980} G.A.~Alekseev and V.A.~Belinski, Sov.Phys.JETP {\bf 51} (4), 655 - 662 (1980)
\bibitem{Figueras-Jamsin-Rocha-Virmani:2010} Pau Figueras, Ella Jamsin, Jorge V. Rocha, Amitabh Virmani, Class.Quant.Grav. {\bf 27}, 135011 (2010); arXiv:0912.3199v1 [hep-th]
\bibitem{Alekseev:2009} G.A.Alekseev, Phys.Rev. {\bf D80}, 041901(R) (2009).
\bibitem{Alekseev:1985} G.A. Alekseev, {\it Sov.\ Phys.\ Dokl.} {\bf 30}, 565 (1985).
\bibitem{Alekseev:1988} G.A. Alekseev, Proc. Steklov Math. Inst., Providence, RI: American Math. Soc., {\bf 3}, 215 (1988).
\bibitem{Alekseev:2005b} G.A. Alekseev, Theor. Math. Phys. {\bf 143}(2), 720-740 (2005); gr-qc/0503043.
\bibitem{Gakhov:1977} F.D.~Gakhov, Boundary Problems [in Russian] (3rd ed.), Nauka, Moscow (1977); English transl. prev. ed.: Boundary Value Problems, Pergamon, Oxford (1966).
\bibitem{Cvetic-Youm:1996} M.~Cveti$\check{\text{c}}$ and D.~Youm, Nucl. Phys. {\bf B 476}, 118 (1996); hep-th/9603100].


\end{thebibliography}
\end{document}